\documentclass[aps,prl,floatfix,showpacs,twocolumn,10pt]{revtex4-2}

\usepackage{graphicx}
\usepackage{amsmath, amsfonts, amssymb, bm}
\usepackage{braket}

\usepackage{amstext}
\usepackage{units}
\usepackage{mathrsfs}
\usepackage{mathtools}

\usepackage{color}
\newcommand{\corr}[1]{{\color{black}#1}}

\begin{document}
\title{Control of nonlinear Compton scattering in a squeezed vacuum}
\author{Antonino~Di Piazza}
\email{a.dipiazza@rochester.edu}
\affiliation{Department of Physics and Astronomy, University of Rochester, Rochester, New York 14627, USA}
\affiliation{Laboratory for Laser Energetics, University of Rochester, Rochester, New York 14623, USA}
\author{Kenan Qu}
\affiliation{Department of Astrophysical Sciences, Princeton University, Princeton, New Jersey 08544, USA}

\begin{abstract}
Electromagnetic radiation by accelerated charges is a fundamental process in physics. Here, we introduce a quantum-optical framework for controlling the emission of radiation of an electron in an intense laser field via squeezed vacuum states. By engineering the quantum fluctuations of the emission modes, we demonstrate that the probability of nonlinear Compton scattering can be significantly enhanced or suppressed through tunable squeezing amplitude and angle. We show numerically that our predictions are experimentally accessible with current squeezing technologies, establishing a new paradigm for quantum control in high-intensity light-matter interactions.
\end{abstract}

\maketitle
The emission of electromagnetic radiation by accelerated charges is a basic process in physics, with applications ranging from synchrotron light sources and astrophysical phenomena to particle accelerators.

If electric charges are accelerated by an electromagnetic field, the process of radiation is described by QED \cite{Landau_b_4_1982,Itzykson_b_1980} and it has a counterpart within classical electrodynamics \cite{Jackson_b_1975,Landau_b_2_1975}. Generally speaking, the field amplitude (angular frequency) scale which determines the importance of quantum effects is given by $F_{cr}=m^2c^3/(\hbar|e|)=1.3\times 10^{16}\;\text{V/cm}=4.4\times 10^{13}\;\text{G}$ ($\omega_C=mc^2/\hbar=7.8\times 10^{20}\;\text{rad/s}$): If an electron experiences in its rest frame a field with amplitude (angular frequency) of the order of or larger than  $F_{cr}$ ($\omega_C$) a full quantum treatment is necessary \cite{Landau_b_4_1982,Dittrich_b_1985,Fradkin_b_1991,Baier_b_1998}. Here, $e<0$ and $m$ are the electron charge and mass, respectively.

Another parameter can be identified, which controls whether nonlinear effects in the electromagnetic field amplitude, i.e., multiphoton effects, are important in the process of radiation. In the case of a laser field, it is given by $\xi_0=|e|E_0/(m\omega_0 c)$, where $E_0$ is the electric field amplitude of the wave and $\omega_0$ its central angular frequency \cite{Ritus_1985,Di_Piazza_2012,Blackburn_2020,Gonoskov_2022,Fedotov_2023}. The threshold condition $\xi_0\sim 1$ for nonlinear effects to become important is already exceeded by several existing high-power lasers \cite{Apollon,CLF,CoReLS,ELI,ZEUS}, whereas future multi-petawatt lasers aim at $\xi_0\gtrsim 100$ \cite{NSF_OPAL,SEL,Vulcan_20-20}. Recent advances in plasma technologies~\cite{Malkin_prl1999, Fisch2003,KQ_prl2017,Zaim_2024} have the potential to reach field strengths near $F_{cr}$, and to generate high-intensity coherent radiation extending into x-rays~\cite{Vincenti_2019,Malkin_4wm2020, Malkin_6wm2023,Zaim_2024}. 

The availability of high-intensity laser radiation as well as of ultrarelativistic electron beams \cite{Gonsalves_2019} has allowed for testing experimentally a new regime of QED, the strong-field QED regime, where both nonlinear and quantum effects influence the electron dynamics \cite{Cole_2018,Poder_2018,Mirzaie_2024,Los_2024}. The properties of radiation in strong-field QED depend on the characteristics of the laser field such as its intensity, pulse shape, or polarization, which have been exploited as control parameters~\cite{Di_Piazza_2012,Fedotov_2023}. In this sense, active control over quantum radiation processes has remained fundamentally classical, i.e., by manipulating the macroscopic properties of the driving laser field.

Recent theoretical works~\cite{Khalaf_2023, np_Even2023, TzurPRR2024, TheidelPRX2024} have begun exploring radiation in quantum light states. However, even in this quantum framework, the modes into which photons are emitted have been treated as an unstructured entity. The question of how engineering the quantum properties of the emission modes themselves could affect quantum radiation in the high-intensity regime has remained theoretically unexplored in the literature and experimentally unrealized.

In this Letter, we introduce a theoretical approach for quantum control of strong-field QED emission by manipulating vacuum fluctuations with quantum squeezing. Unlike prior methods using classical field control, our theoretical approach shows that squeezed vacuum states can directly tailor the quantum emission probability of a photon by an electron in an intense laser field (nonlinear Compton scattering). This paradigm shift moves control from classical mechanisms to a fully quantum approach, in which both the field driving the control and the emission process are inherently quantum. Our theoretical analysis reveals that squeezed vacuum states can either enhance or suppress nonlinear Compton emission probabilities by more than one order of magnitude using experimentally achievable squeezing levels of $10$-$15$ dB~\cite{Kashiwazaki_2023, PRL_15db_2016}. Crucially and unlike in Ref. \cite{Khalaf_2023}, the control relies sensitively on the squeezing angle $\theta_0$, providing precise tunability between suppression ($\theta_0=0$) and enhancement ($\theta_0=\pi$) regimes.

The highest achieved squeezing levels have been exploited to enhance the sensitivity of gravitational-wave detectors \cite{PRL_15db_2016}. A related mechanism to enhance the sensitivity of detection of axions has been put forward in Ref. \cite{Ikeda_2025}. In Ref. \cite{Engineering_prx2017} it was shown how a squeezed vacuum state allows for engineering interactions between electric dipoles. Recent studies propose novel schemes utilizing fully ionized plasmas, potentially enabling squeezing levels from $20$ dB~\cite{Qu_PRE_entangle24} to $40$ dB~\cite{Qu_2025}. Below, we use units with $\epsilon_0=\hbar=c=1$ and the metric tensor $\eta_{\mu\nu}=\text{diag}(+1,-1,-1,-1)$ (the four-dimensional scalar product of two four-vectors $a^{\mu}$ and $b^{\mu}$ is indicated as $(ab)=\eta_{\mu\nu}a^{\mu}b^{\nu}$).

Strong-field QED refers to QED processes occurring in the presence of sufficiently intense background electromagnetic fields that their effects have to be taken into account exactly. This is achieved by quantizing the Dirac field $\psi(x)$, describing electrons and positrons in the presence of the background field (Furry picture)~\cite{Landau_b_4_1982,Fradkin_b_1991,Di_Piazza_2012,Fedotov_2023} whereas the radiation field $A^{\mu}(x)$ is quantized as in vacuum \cite{Itzykson_b_1980}, i.e.,
\begin{equation}
\label{A}
A^{\mu}(x)=\sum_{r=0}^3\int(d^3\bm{k})[a_r(\bm{k})e^{-i(kx)}e^{\mu}_r(\bm{k})+\text{H.c.}],
\end{equation}
where $(d^3\bm{k})=d^3\bm{k}/[(2\pi)^32\omega_k]$, with $k^{\mu}=(\omega_k,\bm{k})$, where $a_r(\bm{k})$ and $a^{\dag}_r(\bm{k})$ are the annihilation and creation operators of a photon with momentum $\bm{k}$ and polarization four-vector $e^{\mu}_r(\bm{k})$, satisfying the commutation relations $[a_r(\bm{k}),a^{\dag}_{r'}(\bm{k}')]=-\eta_{rr}2\omega_k\delta_{r,r'}(2\pi)^3\delta(\bm{k}-\bm{k}')$ and $[a_r(\bm{k}),a_{r'}(\bm{k}')]=0$, and where H.c. stands for Hermitian conjugate. Background-field photons can only be in the transverse modes corresponding to $r=1,2$ and thus we introduce the indices $a=0,3$ and $j=1,2$.

If the background field $\mathcal{A}^{\mu}(x)$ has the form
\begin{equation}
\label{A_B}
\mathcal{A}^{\mu}(x)=\sum_{j=1}^2\int(d^3\bm{k})[b_j(\bm{k})e^{-i(kx)}e^{\mu}_j(\bm{k})+\text{c.c.}],
\end{equation}
with $b_j(\bm{k})$ being complex functions and c.c. standing for complex conjugate, it was shown \cite{Fradkin_b_1991} that the Furry-picture approach is equivalent to the conventional ``vacuum'' interaction picture but with the vacuum state $\ket{0}$ being replaced by the coherent state $\ket{B}=D(b)\ket{0}$, where $D(b)$ is the so-called displacement operator \cite{Mandel_b_2013,Agarwal_b_2013}, defined in Eq. (\ref{D}). A proof is reported for completeness in Supplemental Material \cite{SM}. 

\begin{figure}
\begin{center}
\includegraphics[width=0.8\columnwidth]{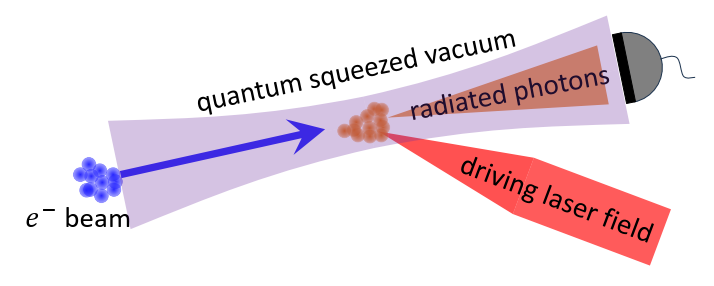}
\end{center}
\vspace*{-8mm}
\caption{Schematics of the considered setup.}
\label{Schematics}
\end{figure}

Our aim is to formulate strong-field QED in the presence of an intense coherent state and in a squeezed vacuum. As sketched in Fig.~\ref{Schematics}, we consider replacing the standard quantum vacuum with a quantum squeezed vacuum in the region where an electron beam collides with an intense driving laser field.
Thus, analogously as above, we replace the vacuum state with the displaced squeezed state $\ket{BZ}=D(b)S(z)\ket{0}$, where $S(z)$ 
is the so-called squeezing operator \cite{Mandel_b_2013,Agarwal_b_2013}, defined in Eq. (\ref{S}).

By using the known properties of the displacement and squeezing operators (see Eqs. (\ref{D_prop}) and (\ref{S_prop})), one can show that $S^{\dag}(z)D^{\dag}(b)A^{\mu}(x)D(b)S(z)=S^{\dag}(z)A^{\mu}(x)S(z)+\mathcal{A}^{\mu}(x)=A^{\mu}_Z(x)+\mathcal{A}^{\mu}(x)$, where
\begin{equation}
\label{A_Z}
A^{\mu}_Z(x)=\sum_{r=0}^3\int(d^3\bm{k})[a_r(\bm{k})E^{\mu}_j(x,\bm{k})+\text{H.c.}],
\end{equation}
with $E^{\mu}_a(x,\bm{k})=e^{\mu}_a(\bm{k})\exp(-i(kx))$ and
\begin{equation}
\label{E_j}
\begin{split}
E^{\mu}_j(x,\bm{k})&=\cosh(\zeta_j(\bm{k}))e^{-i(kx)}e^{\mu}_j(\bm{k})\\
&\quad-\sinh(\zeta_j(\bm{k}))e^{-i\theta_j(\bm{k})}e^{i(kx)}e^{\mu\,*}_j(\bm{k}).
\end{split}
\end{equation}
By repeating the reasoning in the purely coherent case \cite{SM}, we conclude that strong-field QED in a squeezed vacuum can be formulated by quantizing the Dirac field in the Furry picture and the electromagnetic field according to Eq. (\ref{A_Z}). A few remarks are in order: 1) The present method allows one to take into account a background field with pronounced quantum features, like the squeezed field; 2) The function multiplying the annihilation (creation) operator $a_j(\bm{k})$ ($a^{\dag}_j(\bm{k})$) also contains a term proportional to $\exp(i(kx))$ ($\exp(-i(kx))$), which is qualitatively different from the case without squeezing; 3) The case of processes occurring in the squeezed coherent state $\ket{ZB}=S(z)D(b)\ket{0}$ can be worked out analogously by exploiting the commutation relation between the operators $S(z)$ and $D(b)$ \cite{Mandel_b_2013}.

The above procedure is completely general and it offers a wide flexibility in the choice of the coherent and the squeezed modes. However, in order to gain more physical insight on the novelties brought about by the squeezing, we focus our attention onto the case of a background plane-wave field, whose corresponding ``dressed'' electron states have been determined analytically and are known as Volkov states \cite{Volkov_1935,Landau_b_4_1982}. We assume that the plane wave propagates along the positive $z$ direction and that it is polarized along the $x$ direction, corresponding to $j=1$, i.e., $b_j(\bm{k})\to\delta_{j,1}(2\pi)^2\delta(k_x)\delta(k_y)H(k_z)b(k_z)$, where $H(\cdot)$ is the Heaviside step function. Thus, we introduce the four-dimensional quantities $n^{\mu}=(1,0,0,1)$, $a^{\mu}_1=(0,1,0,0)$, and $a^{\mu}_2=(0,0,1,0)$ such that in the Lorenz gauge one can write $\mathcal{A}^{\mu}(\phi)=(0,\bm{\mathcal{A}}(\phi))=A_0f(\phi)a^{\mu}_1$, where $A_0>0$ is a constant describing the amplitude of the field ($\xi_0=|e|A_0/m$), and where $A_0f(\phi)=\int_0^{\infty}\frac{d\omega}{2\pi}\frac{1}{2\omega}[b(\omega)e^{-i\omega\phi}+\text{c.c.}]$, with $\phi=(nx)=t-z$.

At this point the computation of the probability of nonlinear Compton scattering follows exactly the standard procedure, which can be found in the literature (see the reviews \cite{Di_Piazza_2012,Fedotov_2023} and the references therein) but, in the case of a photon emitted with four-momentum $k^{\mu}=(\omega_k,\bm{k})$ and polarization $j$, with the replacement $\exp(i(kx))e^{\mu\,*}_j(\bm{k})\to E_j^{\mu\,*}(x,\bm{k})$. Concerning the incoming electron, we assume that it is described by the normalized wave packet $\Psi_s(x)=\int(d^3\bm{p})\rho(\bm{p})U_s(x;\bm{p})$, where $(d^3\bm{p})=d^3\bm{p}/[(2\pi)^32\varepsilon_p]$, with $\varepsilon_p=\sqrt{m^2+\bm{p}^2}$, where $\rho(\bm{p})$ represents the momentum distribution and $U_s(x;\bm{p})$ is the positive-energy Volkov state with spin quantum number $s$ and momentum $\bm{p}$ outside the field, and normalized as in Ref. \cite{Di_Piazza_2018_d}. If the measured state of the final electron is characterized by the momentum $\bm{p}'$ and the spin quantum number $s'$, we need to compute the transition matrix element from the state $S(z)\ket{0,e_{\Psi_s}}$ to the state $S(z)\ket{\gamma_{\bm{k},j}, e_{\bm{p}',s'}}$ (see also the Appendix B for additional details on these states). By accounting for the field $A_Z^{\mu}(x)$ perturbatively, the leading-order $S$-matrix amplitude $S_j(\bm{k})$ of the process can be written as
\begin{align}
\label{S_fi}
S_j(\bm{k})&=\int(d^3\bm{p})\rho(\bm{p})[(2\pi)^3\delta^3_{\perp,-}(\bm{p}'+\bm{k}-\bm{p})iM_{j,+}(\bm{k},\bm{p}) \nonumber\\
&\quad+(2\pi)^3\delta^3_{\perp,-}(\bm{p}'-\bm{k}-\bm{p})iM_{j,-}(\bm{k},\bm{p})],
\end{align}
where the delta functions correspond to the conservation of the components $P_x=-(a_1P)$, $P_y=-(a_2P)$, and $P_-=(nP)$ of the total four-momentum $P^{\mu}$, and where
\begin{widetext}
\begin{align}
M_{j,\pm}(\bm{k},\bm{p})&=\mp e\frac{e^{\zeta_j(\bm{k})}\pm e^{-\zeta_j(\bm{k})}}{2} e^{i\frac{1\mp 1}{2}\theta_j(\bm{k})}\int d\phi\,e^{i\Phi_{\pm}}\bar{u}_{s'}(\bm{p}')\left[1-e\frac{\hat{n}\hat{\mathcal{A}}(\phi)}{2p'_-}\right]\hat{e}_j(\bm{k})\left[1+e\frac{\hat{n}\hat{\mathcal{A}}(\phi)}{2p_-}\right]u_s(\bm{p})\bigg|_{\substack{
p'_-=p_-\mp k_-\\
\bm{p}'_{\perp}=\bm{p}_{\perp}\mp\bm{k}_{\perp}}},\\
\Phi_{\pm}&=\pm\frac{k_-m^2}{2p_-(p_-\mp k_-)}\int_0^{\phi}d\phi'[1+\bm{\pi}_{\perp}^2(\phi')],\quad \bm{\pi}_{\perp}(\phi)=\frac{\bm{p}_{\perp}}{m}-\frac{p_-}{k_-}\frac{\bm{k}_{\perp}}{m}-e\frac{\bm{\mathcal{A}}_{\perp}(\phi)}{m},
\end{align}
\end{widetext}
with $u_s(\bm{p})$ being the positive-energy constant spinor \cite{Landau_b_4_1982} and with $e^{\mu}_j(\bm{k})$ being chosen real and equal to $a^{\mu}_j-n^{\mu}(ka_j)/k_-$ (the ``hat'' notation indicates the contraction of a four-vector with the Dirac matrices $\gamma^{\mu}$). Note that the process features two interfering paths, one, corresponding to the amplitude $M_{j,+}(\bm{k},\bm{p})$, in which the electron exclusively emits a photon and one, corresponding to the amplitude $M_{j,-}(\bm{k},\bm{p})$, in which the electron also absorbs from the squeezed state two photons with the same four-momentum of the emitted one.

Before computing the corresponding emission probability starting from $|S_j(\bm{k})|^2$, we note that squeezing is most easily realized experimentally in the optical regime. By considering a relativistic electron, whose momentum distribution is well peaked around a momentum $\bm{p}_0$ such that $\bm{p}_0=-p_0\bm{z}$, and a plane wave with $\xi_0\sim 1$ for which nonlinear effects will already be significant, most of the photons will be emitted within a cone centered along $\bm{p}_0$ and of angular aperture of the order of $1/\gamma_0$, with $\gamma_0=\varepsilon_0/m=\sqrt{1+(p_0/m)^2}$. Since under these conditions, the typical emitted frequencies are $\omega\sim 4\gamma_0^2\omega_0$, if we assume that $\gamma_0\sim 10$, i.e. $\varepsilon_0\approx 5\;\text{MeV}$, then for $\omega$ to be in the optical regime, $\omega_0$ should lie in the terahertz (THz) range. Indeed, THz radiation $\xi_0\gtrsim 1$ has been already produced experimentally \cite{Ying_2024,Ying_2025} (see also the experimental proposals \cite{Sheng_2005,Chen_2015,Lu_2025}). Correspondingly, the spatial focusing of the THz radiation can be ignored here as electron beams produced by optical laser fields have typically much smaller transverse areas. Finally, under such experimentally realistic conditions the emitted photon momentum can be safely neglected as compared with the electron momentum, which will simplify the formulas significantly \footnote{A qualitatively new feature arises in the case of large recoil: Whereas without squeezed field the spectrum vanishes in the limit $k_-\to p_-$ due to light-cone energy conservation, this is not the case with the squeezed field precisely due to the new contribution to the transition amplitude proportional to $M_{j,-}(\bm{k},\bm{p})$.}. The width of the wave packet of an electron accelerated to MeV energies is likely to be much larger than the emitted (optical) photon energy (see Ref. \cite{Popov_2004} for the case of a free electron produced via atomic ionization) and the probability $dP_j(\bm{k})$ takes the relatively simple form
\begin{equation}
\begin{split}
&dP_j(\bm{k})=\frac{1}{8\omega_k}\frac{d^3\bm{k}}{(2\pi)^3}\int(d^3\bm{p})\frac{|\rho(\bm{p})|^2}{p_-^2}[1+2\sinh^2(\zeta_j(\bm{k}))\\
&\quad-2\sinh(\zeta_j(\bm{k}))\cosh(\zeta_j(\bm{k}))\cos(\theta_j(\bm{k}))]|M_{j,0}(\bm{k},\bm{p})|^2,
\end{split}
\end{equation}
where $M_{j,0}(\bm{k},\bm{p})$ is the known nonlinear Compton scattering amplitude without squeezing and recoil (see, e.g., Ref. \cite{Di_Piazza_2018}), which can also be obtained from $M_{j,\pm}(\bm{k},\bm{p})$ by setting $\zeta_j(\bm{k})=0$ and neglecting the recoil, i.e., by setting $\bm{p}'\approx \bm{p}$:
\begin{align}
M_{j,0}(\bm{k},\bm{p})&=2e\delta_{ss'}\int d\phi\,e^{i\Phi_0}(P(\phi)e_j(\bm{k})),\\
\Phi_0&=k_-\frac{m^2}{2p_-^2}\int_0^{\phi}d\phi'[1+\bm{\pi}_{\perp}^2(\phi')],
\end{align}
where we have also used Eqs. (\ref{Identity_1})-(\ref{Identity_2}). Note that although the quantum effect of photon recoil is negligible here, the quantum treatment based on strong-field QED is necessary to include the nonlinear effects in the THz laser field ($\xi_0\sim 1$), as well as the effects of squeezing and of the electron wave packet. Finally, if the squeezing is implemented along the polarization $j=1$, only the corresponding emission probability is affected. By indicating as $d\mathcal{P}_1(\bm{k})$ the differential probability averaged/summed over the initial/final electron spins, we obtain $d\mathcal{P}_1(\bm{k})=[1+F_1(\bm{k})]d\mathcal{P}^{(0)}_1(\bm{k})$, where
\begin{equation}
\begin{split}
F_1(\bm{k})&=2\sinh(\zeta_1(\bm{k}))\left[e^{\zeta_1(\bm{k})}\sin^2\left(\frac{\theta_1(\bm{k})}{2}\right)\right.\\
&\quad\left.-e^{-\zeta_1(\bm{k})}\cos^2\left(\frac{\theta_1(\bm{k})}{2}\right)\right],
\end{split}
\end{equation}
and where
\begin{equation}
\label{Prob_spec}
d\mathcal{P}^{(0)}_1(\bm{k})=\frac{\alpha m^2}{\omega_k}\frac{d^3\bm{k}}{4\pi^2}\int(d^3\bm{p})\frac{|\rho(\bm{p})|^2}{p_-^2}\left|\int d\phi\, e^{i\Phi_0}\pi_1(\phi)\right|^2
\end{equation}
is the corresponding differential probability without squeezing, with $\alpha=e^2/4\pi\approx 1/137$. Thus, the effects of squeezing are encoded in the function $F_1(\bm{k})$. In particular and unlike the results in Ref. \cite{Khalaf_2023}, we find that the emission probability does depend on the squeezing angle, which can qualitatively affect the correction: if $\theta_1(\bm{k})=0$ ($\theta_1(\bm{k})=\pi$) then the correction is negative (positive) and proportional to $1-\exp[-2\zeta_1(\bm{k})]$ ($\exp[2\zeta_1(\bm{k})]-1$). This reflects the corresponding suppression/enhancement effect in the effective ``photon wave function'' $E^{\mu}_j(x,\bm{k})$ in the case of small recoil.

\begin{figure}
\begin{center}
\includegraphics[width=0.8\columnwidth]{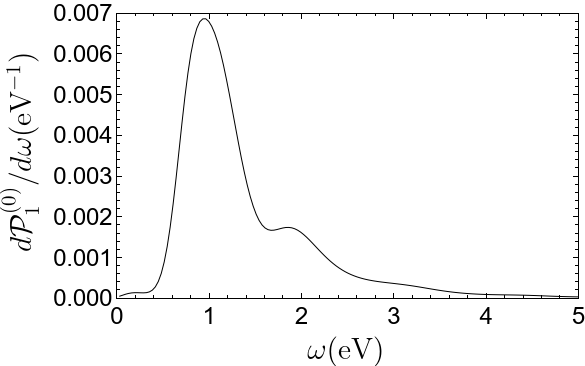}
\end{center}
\vspace*{-8mm}
\caption{Probability spectrum for numerical parameters given in the text.}
\label{Spectrum}
\end{figure}

Now, we consider a THz laser field with central wavelength of $200\;\text{$\mu$m}$, Gaussian envelope with a full-width half maximum of $\tau=1\;\text{ps}$, and peak intensity $3\times 10^{13}\;\text{W/cm$^2$}$ (corresponding to $\xi_0=1$) colliding with an electron wave packet with $\varepsilon_0=5\;\text{MeV}$. In such an intense THz field, an extension of the wave packet in momentum space of the order of $1\;\text{eV}$ does not alter the probability spectrum (\ref{Prob_spec}), which has been plotted in Fig. \ref{Spectrum} for an electron with momentum $\bm{p}_0$ and by integrating it over a solid angle $\delta\Omega$ corresponding to a narrow cone centered along the negative $z$-axis and of aperture $m\xi_0/\varepsilon_0\sim 0.1\;\text{rad}$ \footnote{The polarization state of the emitted photon has been chosen to depend on the emission angle. However, for the chosen aperture cone, the polarization is approximately constant and along the $x$ direction.}.

Correspondingly, the squeezed modes are centered around the wave-vector $\bm{k}_{\text{sq}}=-\omega_{\text{sq}}\bm{z}$, with $\omega_{\text{sq}}\approx 1\;\text{eV}$ to match the maximum in the spectrum in Fig. \ref{Spectrum} and with a uniform angular distribution within a cone of solid-angle aperture $\Delta\Omega>\delta\Omega$. Thus, the function $F_1(\bm{k})$ depends only on $\omega$ within $\delta\Omega$ and $d\mathcal{P}_1/d\omega=[1+F(\omega)]d\mathcal{P}^{(0)}_1/d\omega$, with $F(\omega)=F_1(\bm{k})$ there. The angular frequency distribution is given by a standard Lorentzian shape \cite{PRL_15db_2016}:
\begin{equation}
\zeta_1(\bm{k})=\frac{\zeta_0}{1+(\omega-\omega_{\text{sq}})^2/\Gamma^2},
\end{equation}
with $\zeta_0$ and $\Gamma$ being positive constants. Instead, the distribution of the squeezing angle is assumed to be uniform: $\theta_1(\bm{k})=\theta_0$. 

\begin{figure}
\begin{center}
\includegraphics[width=0.8\columnwidth]{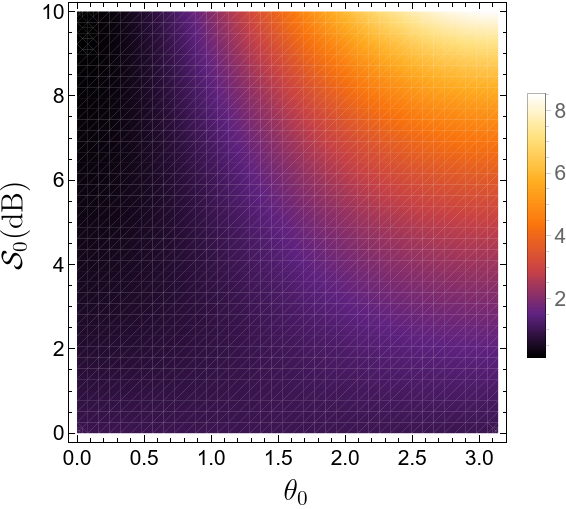}
\end{center}
\vspace*{-8mm}
\caption{Probability control factor $\mathcal{C}(\zeta_0,\theta_0)$ as a function of $\mathcal{S}_0(\text{dB})=20\,\zeta_0/\log(10)$ and of $\theta_0$.}
\label{Enhancement}
\end{figure}

According to the above analysis, the observable is the polarized emitted radiation within the narrow solid angle $\delta\Omega$ centered along the negative $z$ direction and in a frequency window $\Gamma$ centered around $\omega_{\text{sq}}$ (see also Fig. \ref{Schematics}). We indicate as $|\Delta^3\bm{k}_{\text{sq}}|\approx \omega_{\text{sq}}^2 \delta\Omega\times\Gamma$ this volume in momentum space. 
Since the frequency width $\Gamma$ lies in the THz range, the value of the main peak in Fig. \ref{Spectrum} is essentially constant within $\Gamma$ and the effects of the squeezing can be estimated quite universally via the average function
\begin{equation}
\begin{split}
&\mathcal{F}(\zeta_0,\theta_0)=\frac{1}{|\Delta^3\bm{k}_{\text{sq}}|}\int_{|\Delta^3\bm{k}_{\text{sq}}|}d^3\bm{k}\,F_1(\bm{k})\\
&\quad\approx\sin^2\left(\frac{\theta_0}{2}\right)\frac{1}{2}\int_{-1}^1d\eta\left(e^{\frac{2\zeta_0}{1+\eta^2/4}}-1\right)\\
&\qquad-\cos^2\left(\frac{\theta_0}{2}\right)\frac{1}{2}\int_{-1}^1d\eta\left(1-e^{-\frac{2\zeta_0}{1+\eta^2/4}}\right).
\end{split}
\end{equation}
In Fig. \ref{Enhancement} we plot the probability control factor $\mathcal{C}(\zeta_0,\theta_0)=1+\mathcal{F}(\zeta_0,\theta_0)$ as a function of the squeezing amplitude $\mathcal{S}_0(\text{dB})=20\,\zeta_0/\log(10)$ and of the squeezing angle $\theta_0$ \corr{(recall that the spectrum including squeezing is obtained by multiplying $d\mathcal{P}^{(0)}_1/d\omega$ by $\mathcal{C}(\zeta_0,\theta_0)$)}.
The figure shows how the squeezing effect crucially depends on the squeezing angle. Indeed, for $\theta_0\ll 1$ ($0<\pi-\theta_0\ll 1$) and for $\mathcal{S}_0\approx 10\;\text{dB}$ squeezing can already significantly suppress (enhance) the probability of emitting a photon in the squeezed mode by a factor of \corr{$0.12$ ($8.5$)}. The values of $\mathcal{S}_0$ in Fig. \ref{Enhancement} are well within the reach of present technology. For $\mathcal{S}_0=15\;\text{dB}$ \cite{PRL_15db_2016} the suppression (enhancement) factor at $\theta_0=0$ ($\theta_0=\pi$) would even be about \corr{$0.042$ ($25$)}. For the sake of completeness, it should be recalled that the photons of the squeezed field also propagate along the direction of observation and with the same energy and polarization. \corr{However, the average number $N_{\text{sq}}$ of photons in the squeezed state $S(z)\ket{0}$ within the interaction volume $V$ and with momentum between $\bm{k}$ and $\bm{k}+\delta^3\bm{k}$ and with polarization $j$ is given by $n_j(\bm{k})V\delta^3\bm{k}/(2\pi)^3$, where $n_j(\bm{k})=\sinh^2(\zeta_j(\bm{k}))$ \cite{Mandel_b_2013,Agarwal_b_2013}. In our case, we can estimate $V\sim\pi\sigma_e^2\tau$, where $\sigma_e\sim 5\;\text{$\mu$m}$ is the transverse size of the electron beam, and $|\delta^3\bm{k}|\sim |\Delta^3\bm{k}_{\text{sq}}|$, with $\delta\Omega\sim 1/\gamma_0^2$. The result is that $N_{\text{sq}}\lesssim 10$} for $\mathcal{S}=15\;\text{dB}$ and thus completely negligible as compared to the expected number of photons emitted without squeezing by a typical electron beam with 100 pC charge \cite{Gonsalves_2019}, which is of the order of $d\mathcal{P}^{(0)}_1(1\;\text{eV})/d\omega\times \Gamma\times 10^9\approx 3\times 10^4$ for $\Gamma= 1\;\text{THz}$ $(\approx 4\;\text{meV}$) (see Fig. \ref{Spectrum}). 
\corr{Although the bandwidth $\Gamma \approx 4$ meV is narrow compared to the eV-scale emission spectrum, it is well within the resolution of standard optical spectrometers~\cite{Sennary_2025}.} 

The possibility of controlling the sign of the effect indicates that it is not simply ``stimulated'' emission from an already populated state. Had we considered nonlinear Compton emission of a photon in a state where already $N\gg 1$ photons would be present instead that in a squeezed vacuum, we would have obtained an enhancement of the probability by factor of approximately $N$. This hints that an analogous enhancement would have been obtained in the presence of a coherent field along the emitted photon.

It is also worth observing here that our theoretical approach would not be suitable for too large enhancements because we treat the interaction between the Dirac field and the squeezed field $A^{\mu}_Z(x)$ perturbatively. In order to quantify when such an approach is valid we start from the average number $n_j(\bm{k})$ of squeezed photons given above and estimate the average energy density of the squeezed field as in Ref. \cite{Landau_b_4_1982}. One can then obtain an estimate of the classical nonlinearity parameter $\xi_{0,\text{sq}}$ of this field to be $\xi_{0,\text{sq}}\sim\sqrt{\alpha}(\omega_{\text{sq}}/m)\text{exp}(\zeta_0)$. Thus, the above approach is valid for $\xi_{0,\text{sq}}\ll 1$, which is easily fulfilled for realistic values of $\zeta_0$ ($\sim 1$) if $\omega_{\text{sq}}$ is in the optical range ($\sim 1\;\text{eV}$), as $\sqrt{\alpha}(\omega_{\text{sq}}/m)\sim 10^{-7}$. 

Finally, we observe that \corr{recent experiments have demonstrated squeezing over $7$ THz~\cite{Chen_2022} and even sub-petahertz~\cite{Sennary_2025} bandwidths. While these ultrabroadband sources currently exhibit lower squeezing magnitudes than narrowband sources, there is no fundamental physical barrier to increasing the squeezing level as the technology matures. Furthermore,} a series of recent experiments has begun to explore the extension of squeezing into shorter wavelengths via high-harmonic generation~\cite{TheidelPRX2024, TzurPRR2024}. Also, the realization of squeezed vacuum states in the x-ray regime is becoming increasingly plausible~\cite{Qu_PRE_entangle24,Qu_2025}. The above results would also be valid in that frequency range because the effects of recoil can still be safely neglected for an MeV-electron. Moreover, this latter setup would have the advantage that the intense laser field could be optical.

In conclusion, we have shown that the probability of nonlinear Compton scattering can be controlled by squeezing the vacuum at the expected emitted modes. Apart from the squeezing amplitude, an important steering parameter is the squeezing angle, which crucially controls whether squeezing enhances or suppresses the probability. Our numerical results show that the presented predictions can in principle be tested experimentally by colliding a moderately relativistic electron beam with either an intense THz laser field and squeezing in the optical regime or an intense optical laser field and squeezing in the x-ray domain. The control is sensitive to the photon polarization, such that optical or x-ray beams with controlled polarization can be in principle produced, with multiple possible applications spanning from light-matter interaction to quantum optics and quantum information.

\begin{acknowledgments}
A.D.P. is partially supported by the U.S. National Science Foundation Mid-scale Research Infrastructure Program under Award No. PHY-2329970. K.Q. is supported by NNSA Grant No. DE-NA0004167

This material is based upon work supported by the U.S. Department of Energy [National Nuclear Security Administration] University of Rochester ``National Inertial Confinement Fusion Program'' under Award Number DE-NA0004144.

This report was prepared as an account of work sponsored by an agency of the United States Government. Neither the United States Government nor any agency thereof, nor any of their employees, makes any warranty, express or implied, or assumes any legal liability or responsibility for the accuracy, completeness, or usefulness of any information, apparatus, product, or process disclosed, or represents that its use would not infringe privately owned rights. Reference herein to any specific commercial product, process, or service by trade name, trademark, manufacturer, or otherwise does not necessarily constitute or imply its endorsement, recommendation, or favoring by the United States Government or any agency thereof. The views and opinions of authors expressed herein do not necessarily state or reflect those of the United States Government or any agency thereof.

A.D.P. gratefully acknowledges insightful discussions with K. Z. Hatsagortsyan, C. Kleine, G. T. Landi, and J. Luiten.

\end{acknowledgments}
\paragraph{Appendix A: Displacement and squeezing operators}
The displacement and the squeezing operators are fundamental unitary operators in quantum optics \cite{Mandel_b_2013,Agarwal_b_2013}. By introducing the complex functions $b_j(\bm{q})$ and $z_j(\bm{q})$, the displacement operator $D(b)$ and the squeezing operator $S(z)$ are defined as \cite{Mandel_b_2013,Agarwal_b_2013}
\begin{align}
\label{D}
D(b)&=\exp\left\{\sum_{j=1}^2\int(d^3\bm{q})[b_j(\bm{q})a^{\dag}_j(\bm{q})-\text{H.c.}]\right\},\\
\label{S}
S(z)&=\exp\left\{\frac{1}{2}\sum_{j=1}^2\int(d^3\bm{q})[z^*_j(\bm{q})a^2_j(\bm{q})-\text{H.c.}]\right\}.
\end{align}
By using the commutation relations among photon creation and annihilation operators, the basic properties of the operators $D(b)$ and $S(z)$
\begin{align}
\label{D_prop}
D^{\dag}(b)a_r(\bm{k})D(b)&=a_r(\bm{k})+\sum_{j=1}^2\delta_{rj}b_j(\bm{k}),\\
\label{S_prop}
\begin{split}
S^{\dag}(z)a_r(\bm{k})S(z)&=a_r(\bm{k})+\sum_{j=1}^2\delta_{rj}\Big\{[\cosh(\zeta_j(\bm{k}))-1]\\
&\quad\left.\times a_j(\bm{k})-\sinh(\zeta_j(\bm{k}))e^{i\theta_j(\bm{k})}a^{\dag}_j(\bm{k})\right\}
\end{split}
\end{align}
can be easily proven \cite{Mandel_b_2013,Agarwal_b_2013}, where $z_j(\bm{q})=\zeta_j(\bm{q})\exp(i\theta_j(\bm{q}))$.

\paragraph{Appendix B: Properties of initial and final states}
In this appendix, we describe some of the properties of the quantum states used in our derivation especially to interpret the photon statistics of the final state.

The calculation of the $S$-matrix was performed for a transition from the initial state $S(z)\ket{0,e_{\Psi_s}}$ to the final state $S(z)\ket{\gamma_{\bm{k},j}, e_{\bm{p}',s'}}$. In the discussion below, we ignore the states of the electrons and focus on the states of the photons: $\ket{Z}=S(z)\ket{0}$ and $\ket{Z;\bm{k},j}=S(z)a_j^{\dag}(\bm{k})\ket{0}$, where $\ket{0}$ is the vacuum state normalized to unity: $\braket{0|0}=1$. Because the squeezing operator  $S(z)$ is unitary, it preserves the normalization of the states, i.e., $\braket{Z|Z}=1$ and $\braket{Z;\bm{k},j|Z;\bm{k}',j'}=2\omega_k\delta_{j,j'}(2\pi)^3\delta^3(\bm{k}-\bm{k}')$ \cite{Itzykson_b_1980}. 

While the specific final photon state $S(z)a_j^{\dag}(\bm{k})\ket{0}$ is chosen for convenience of normalization, it is equivalent to a single-photon-added squeezed vacuum state, i.e., to $\ket{\bm{k},j;Z}=a_j^{\dag}(\bm{k})S(z)\ket{0}$, apart from a normalization factor
\begin{equation}
\begin{split}
\ket{\bm{k},j;Z}&=S(z)[\cosh(\zeta_j(\bm{k}))a_j^{\dag}(\bm{k})\\
&\quad-\sinh(\zeta_j(\bm{k}))e^{i\theta_j(\bm{k})}a_j(\bm{k})]\ket{0}\\
&=\cosh(\zeta_j(\bm{k}))\ket{Z;\bm{k},j}.
\end{split}
\end{equation}
Thus, using either state yields the same physical results for the transition probability.

Now, the state $\ket{\bm{k},j;Z}$ is nonclassical and its photon distribution statistics adds complexity to the computation of average photon numbers. For example, the average number of photons in the initial squeezed vacuum state depends on the squeezing magnitude $\zeta_i(\bm{q})$, i.e.,
\begin{equation}
\braket{Z|\mathcal{N}|Z}=V\sum_{i=1}^2\int d^3\bm{q}\,n_i(\bm{q}),
\end{equation}
where $\mathcal{N}=\sum_{i=1}^2\int (d^3\bm{q})a^{\dag}_i(\bm{q})a_i(\bm{q})$ is the number operator~\cite{Mandel_b_2013,Agarwal_b_2013}, $n_i(\bm{q})=\sinh^2(\zeta_i(\bm{q}))$ and $V$ is the quantization volume. 

For the final state $\ket{\bm{k},j;Z}$ and its above-mentioned physical interpretation, we would expect that the average number of photons is something like $1+\braket{Z|\mathcal{N}|Z}$. Instead, a direct calculation provides
\begin{equation}
\begin{split}
\frac{\braket{\bm{k},j;Z|\mathcal{N}|\bm{k},j;Z}}{\braket{\bm{k},j;Z|\bm{k},j;Z}}&=V\sum_{i=1}^2\int d^3\bm{q}\sinh^2(\zeta_i(\bm{q}))\\
&\quad+\cosh(2\zeta_j(\bm{k}))\\
&=1+\braket{Z|\mathcal{N}|Z}+2\sinh^2(\zeta_j(\bm{k})).
\end{split}
\end{equation}
This result can be understood by referring to the fact that the number of photons in the state $\ket{Z}$ is not definite but it has a variance $\sigma^2_{\mathcal{N}}=\braket{Z|\mathcal{N}^2|Z}-\braket{Z|\mathcal{N}|Z}^2=V\sum_{i=1}^2\int d^3\bm{q}\,\sigma^2_i(\bm{q})$, with
\begin{equation}
\sigma^2_i(\bm{q})=2\sinh^2(\zeta_i(\bm{q}))\cosh^2(\zeta_i(\bm{q}))
\end{equation}
such that 
\begin{equation}
\frac{\braket{\bm{k},j;Z|\mathcal{N}|\bm{k},j;Z}}{\braket{\bm{k},j;Z|\bm{k},j;Z}}=1+\braket{Z|\mathcal{N}|Z}+\frac{\sigma^2_j(\bm{k})}{1+n_j(\bm{k})}.
\end{equation}
That this is indeed the correct result can be easily proved by considering a single-mode harmonic oscillator with $n$-particle state $\ket{n}$, annihilation (creation) operator $a$ ($a^{\dag}$), and number operator $N=a^{\dag}a$. In fact, if we consider the state $\ket{c}=\sum_{n=1}^{\infty}c_n\ket{n}$, with $\sum_{n=1}^{\infty}|c_n|^2=1$, and the state $\ket{c+1}=a^{\dag}\ket{c}$ with one additional creation operator, it is
\begin{equation}
\frac{\braket{c+1|N|c+1}}{\braket{c+1|c+1}}=1+\braket{c|N|c}+\frac{\sigma^2}{1+\braket{c|N|c}},
\end{equation}
with $\sigma^2=\braket{c|N^2|c}-\braket{c|N|c}^2$.

\paragraph{Appendix C: A useful identity}
Starting from the identity $\bar{u}_{s'}(\bm{p})\gamma^{\mu}u_s(\bm{p})=2p^{\mu}\delta_{ss'}$, it can be shown that \cite{Landau_b_4_1982}
\begin{equation}
\label{Identity_1}
\begin{split}
&\bar{u}_{s'}(\bm{p})\left[1-e\frac{\hat{n}\hat{\mathcal{A}}(\phi)}{2p_-}\right]\gamma^{\mu}\left[1+e\frac{\hat{n}\hat{\mathcal{A}}(\phi)}{2p_-}\right]u_s(\bm{p})\\
&=2P^{\mu}(\phi)\delta_{ss'},
\end{split}
\end{equation}
where
\begin{equation}
\label{Identity_2}
P^{\mu}(\phi)=p^{\mu}-e\mathcal{A}^{\mu}(\phi)+\frac{e(p\mathcal{A}(\phi))}{p_-}n^{\mu}-\frac{e^2\mathcal{A}^2(\phi)}{2p_-}n^{\mu}
\end{equation}
is the classical four-momentum of an electron with initial four-momentum $p^{\mu}$ in the plane wave \cite{Landau_b_2_1975}.

\begin{widetext}
\section*{Supplemental Material: Control of nonlinear Compton scattering via squeezed coherent radiation}
In order to clarify the relation between processes occurring in the presence of a coherent state and the use of the Furry picture, a derivation of the latter is presented, which is equivalent to the original one published in Ref. \cite{Furry_1951}.

The Lagrangian density of strong-field QED in the Lorenz-Feynman gauge is given by \cite{Landau_b_4_1982,Fradkin_b_1991,Di_Piazza_2012,Fedotov_2023}
\begin{equation}
\mathcal{L}=\bar{\psi}(x)(i\gamma^{\mu}\partial_{\mu}-m)\psi(x)-\frac{1}{2}[\partial_{\mu}A_{\nu}(x)][\partial^{\mu}A^{\nu}(x)]-e\bar{\psi}(x)\gamma^{\mu}\psi(x)[A_{\mu}(x)+\mathcal{A}_{\mu}(x)],
\end{equation}
where $\psi(x)$ is the Dirac field, $A^{\mu}(x)$ ($\mathcal{A}^{\mu}(x)$) is the radiation (background) electromagnetic field, and units with $\epsilon_0=\hbar=c=1$ are used (the metric tensor is $\eta_{\mu\nu}=\text{diag}(+1,-1,-1,-1)$). The field  $\bar{\psi}(x)=\psi^{\dag}(x)\gamma^0$ is the Dirac conjugated of $\psi(x)$, with $\gamma^{\mu}$ being the Dirac gamma matrices in the standard representation, and it has been implicitly assumed that the quantization of the electromagnetic field will be carried out within the Gupta-Bleuler formalism to overcome the problems related to the indefinite metric, where any physical state $\ket{\Psi}$ satisfies the relation $\braket{\Psi|\partial_{\mu}A^{\mu}(x)|\Psi}=0$ \cite{Itzykson_b_1980}. Analogously we assume that $\partial_{\mu}\mathcal{A}^{\mu}(x)=0$ and also that $\lim_{t\to-\infty}\mathcal{A}^{\mu}(x)=0$.

The time-evolution of the dynamical fields $\psi(x)$ and $A^{\mu}(x)$ can be described in the standard interaction representation, which uses the first two terms in $\mathcal{L}$ as free Lagrangian density and the last one as interaction Lagrangian density. In this representation, the fields $\psi(x)=\psi_V(x)$ and $A^{\mu}(x)$ satisfy the (operator) equations
\begin{align}
(\gamma^{\mu}i\partial_{\mu}-m)\psi_V(x)&=0,\\
\partial_{\mu}\partial^{\mu}A^{\nu}(x)&=0.
\end{align}
The reason  why we have added the subscript $V$ for ``vacuum'' to the Dirac field in this representation will be clear below.

The spaces of solutions of these equations admit the plane-wave states with definite momentum and spin/polarization quantum numbers as bases such that the operators $\psi_V(x)$ and $A^{\mu}(x)$ can be written as
\begin{align}
\psi_V(x)&=\sum_{s=1}^2\int\frac{d^3\bm{p}}{(2\pi)^3}\frac{1}{2\varepsilon_p}[c_s(\bm{p})e^{-i(px)}u_s(\bm{p})+d^{\dag}_s(\bm{p})e^{i(px)}v_s(\bm{p})],\\
A^{\mu}(x)&=\sum_{r=0}^3\int\frac{d^3\bm{k}}{(2\pi)^3}\frac{1}{2\omega_k}[a_r(\bm{k})e^{-i(kx)}e^{\mu}_r(\bm{k})+a^{\dag}_r(\bm{k})e^{i(kx)}e^{\mu\,*}_r(\bm{k})],
\end{align}
where $p^{\mu}=(\varepsilon_p,\bm{p})$, with $\varepsilon_p=\sqrt{m^2+\bm{p}^2}$, $k^{\mu}=(\omega_k,\bm{k})$, with $\omega_k=\sqrt{\bm{k}^2}$, $u_s(\bm{p})$ and $v_s(\bm{p})$ are the constant positive- and negative-energy spinors, $e^{\mu}_r(\bm{k})$ are the constant polarization four-vectors, and $c_s(\bm{p})$, $c^{\dag}_s(\bm{p})$, $d_s(\bm{p})$, $d^{\dag}_s(\bm{p})$, $a_r(\bm{k})$, and $a^{\dag}_r(\bm{k})$ etc... are the electron, positron, and photon annihilation and creation operators in the states with the corresponding quantum numbers and satisfy the usual (anti-)commutation relations $\{c_s(\bm{p}),c^{\dag}_{s'}(\bm{p}')\}=\{d_s(\bm{p}),d^{\dag}_{s'}(\bm{p}')\}=2\varepsilon_p\delta_{s,s'}(2\pi)^3\delta(\bm{p}-\bm{p}')$, $[a_r(\bm{k}),a^{\dag}_{r'}(\bm{k}')]=-2\omega_k\eta_{rr'}(2\pi)^3\delta(\bm{k}-\bm{k}')$ etc....

Within this representation, the states $\ket{\Psi(t)}_V$ of the system satisfy the equation
\begin{equation}
i\frac{d\ket{\Psi(t)}_V}{dt}=\int d^3\bm{x}\,j_V^{\mu}(x)[A_{\mu}(x)+\mathcal{A}_{\mu}(x)]\ket{\Psi(t)}_V\\
\end{equation}
with $j^{\mu}_V(x)=e\bar{\psi}_V(x)\gamma^{\mu}\psi_V(x)$, and evolve according to the time-evolution operator
\begin{equation}
U_V(t,t')=\mathcal{T}\left(e^{-i\int_{t'}^td\tilde{t}\int d^3\bm{x}j^{\mu}_V(\tilde{x})[A_{\mu}(\tilde{x})+\mathcal{A}_{\mu}(\tilde{x})]}\right),
\end{equation}
where $\mathcal{T}$ is the time-ordering operator, $t\ge t'$, and $\tilde{x}=(\tilde{t},\bm{x})$ (we ignore here the subtleties related to the normal ordering of the interaction Lagrangian density, which are irrelevant for the present discussion, see, e.g., \cite{Itzykson_b_1980}). Thus, if $\ket{i(-\infty)}_V$ and $\ket{f(+\infty)}_V$ represent two states featuring a certain number of electrons, positrons, and photons with certain quantum numbers for $t\to -\infty$ and $t\to+\infty$, respectively, the transition amplitude from $\ket{i(-\infty)}_V$ to $\ket{f(+\infty)}_V$ is given by
\begin{equation}
\prescript{}{V}{\braket{f(+\infty)|S_V|i(-\infty)}}^{}_V,
\end{equation}
where $S_V=U_V(+\infty,-\infty)$ is the scattering operator.

The (equivalent) description of the system within the Furry representation is implemented via the unitary operator $\mathcal{U}(t)$ which satisfies the equation
\begin{equation}
i\frac{d\mathcal{U}(t)}{dt}=-\mathcal{U}(t)\int d^3\bm{x}\,j^{\mu}_V(x)\mathcal{A}_{\mu}(x),
\end{equation}
with the initial condition $\mathcal{U}(t_0)=I$, if it is assumed that the two descriptions coincide at $t=t_0$. In order to write down a formal solution of this equation one needs to distinguish whether $t>t_0$ or $t<t_0$. For example, in the former case, it is
\begin{equation}
\mathcal{U}(t)=\tilde{\mathcal{T}}\left(e^{i\int_{t_0}^td\tilde{t}\int d^3\bm{x}\,j^{\mu}_V(\tilde{x})\mathcal{A}_{\mu}(\tilde{x})}\right),
\end{equation}
where $\tilde{\mathcal{T}}$ is the time-anti-ordering operator. However, the considerations below do not depend on the form of the solution but only on the differential equation satisfied by the operator $\mathcal{U}(t)$.

The state and the operators in the Furry representation are defined as
\begin{align}
\ket{\Psi(t)}_F&=\mathcal{U}(t)\ket{\Psi(t)}_V,\\
\psi_F(x)&=\mathcal{U}(t)\psi_V(x)\mathcal{U}^{-1}(t),\\
A^{\mu}_F(x)&=\mathcal{U}(t)A^{\mu}_V(x)\mathcal{U}^{-1}(t)=A^{\mu}(x).
\end{align}
It is, in fact, straightforward to show that the state $\ket{\Psi(t)}_F$ and the Dirac operator $\psi_F(x)$ satisfy the equations:
\begin{align}
\label{Schroedinger_Furry}
i\frac{d\ket{\Psi(t)}_F}{dt}&=\int d^3\bm{x}\,j^{\mu}_F(x)A_{\mu}(x)\ket{\Psi(t)}_F,\\
\{\gamma^{\mu}[i\partial_{\mu}-e\mathcal{A}_{\mu}(x)]-m\}\psi_F(x)&=0,
\end{align}
with $j^{\mu}_F(x)=e\bar{\psi}_F(x)\gamma^{\mu}\psi_F(x)$. In order to prove the second equation one exploits the fact that the anti-commutation rules for the Dirac field and its conjugated in the Furry representation are the same as in vacuum \cite{Itzykson_b_1980}:
\begin{align}
\{\psi_F(t,\bm{x}),\psi_F(t,\bm{x}')\}&=\{\psi_V(t,\bm{x}),\psi_V(t,\bm{x}')\}=0,\\
\{\psi_F(t,\bm{x}),\bar{\psi}_F(t,\bm{x}')\}&=\{\psi_V(t,\bm{x}),\bar{\psi}_V(t,\bm{x}')\}=\gamma^0\delta^3(\bm{x}-\bm{x}'),\\
\{\bar{\psi}_F(t,\bm{x}),\bar{\psi}_F(t,\bm{x}')\}&=\{\bar{\psi}_V(t,\bm{x}),\bar{\psi}_V(t,\bm{x}')\}=0,
\end{align}
and that
\begin{equation}
[j^{\mu}_F(t,\bm{x}),\psi_F(t,\bm{x}')]=-\{\psi_F(t,\bm{x}'),\bar{\psi}_F(t,\bm{x})\}\gamma^{\mu}\psi_F(x)=-\delta^3(\bm{x}-\bm{x}')\gamma^0\gamma^{\mu}\psi_F(x).
\end{equation}

By introducing the time-evolution operator $U_F(t,t')$ in the Furry representation according to the relation $\ket{\Psi(t)}_F=U_F(t,t')\ket{\Psi(t')}_F$ for $t\ge t'$, it is clear that the identities hold
\begin{align}
U_F(t,t')&=\mathcal{U}(t)U_V(t,t')\mathcal{U}^{-1}(t'),\\
U_F(t,t')&=\mathcal{T}\left(e^{-i\int_{t'}^td\tilde{t}\int d^3\bm{x}\,j^{\mu}_F(\tilde{x})A_{\mu}(\tilde{x})}\right).
\end{align}
The second identity, in particular, can be proved from Eq. (\ref{Schroedinger_Furry}) or by explicitly observing that the operator $U_F(t,t')$ has to satisfy the differential equation
\begin{equation}
\begin{split}
i\frac{\partial U_F(t,t')}{\partial t}&=-\mathcal{U}(t)\int d^3\bm{x}\,j_V^{\mu}(x)\mathcal{A}_{\mu}(x)\mathcal{U}(t)U_V(t,t')\mathcal{U}^{-1}(t')\\
&\quad+\mathcal{U}(t)\int d^3\bm{x}\,j_V^{\mu}(\tilde{x})[A_{\mu}(\tilde{x})+\mathcal{A}_{\mu}(\tilde{x})]U_V(t,t')\mathcal{U}^{-1}(t')\\
&=\int d^3\bm{x}\,j_F^{\mu}(\tilde{x})A_{\mu}(\tilde{x}) U_F(t,t'),
\end{split}
\end{equation}
for $t\ge t'$ with the initial condition $U_F(t',t')=I$.

Finally, the transition matrix element between the initial state  $\ket{i(-\infty)}_F=\mathcal{U}(-\infty)\ket{i(-\infty)}_V$ and the final state $\ket{f(+\infty)}_F=\mathcal{U}(+\infty)\ket{i(+\infty)}_V$, is given by the equation
\begin{equation}
\begin{split}
\prescript{}{F}{\braket{f(+\infty)|S_F|i(-\infty)}}^{}_F&=\prescript{}{F}{\braket{f(+\infty)|U_F(+\infty,-\infty)|i(-\infty)}}^{}_F\\
&=\prescript{}{V}{\braket{f(+\infty)|U_V(+\infty,-\infty)|i(-\infty)}}^{}_V=\prescript{}{V}{\braket{f(+\infty)|S_V|i(-\infty)}}^{}_V,
\end{split}
\end{equation}
where the scattering operator $S_F=U_F(+\infty,-\infty)$ in the Furry representation has been introduced, which further corroborates the equivalence between the two representations.

In this respect, one has to point out that the unitarity of the operator $\mathcal{U}(t)$ is equivalent to the possibility of quantizing the Dirac field in the Furry representation analogously as in the vacuum. This significantly limits the background fields to which the Furry picture can be applied \cite{Fradkin_b_1991}. Specifically the background field must be such that a basis of the solutions of the corresponding Dirac equations can be found and that positive-energy- and negative-energy-like solutions can be identified featuring a finite energy gap at any time $t$ \cite{Fradkin_b_1991}. We worked under this assumption in the main text, which is fulfilled in the presence of a plane-wave background field \cite{Landau_b_4_1982,Fradkin_b_1991}.

Finally, we need to connect the standard interaction picture, with the Dirac field being quantized in vacuum and the interaction Lagrangian density being given by $-e\bar{\psi}(x)\gamma^{\mu}\psi(x)[A_{\mu}(x)+\mathcal{A}_{\mu}(x)]$ with the same interaction picture without the background field $\mathcal{A}^{\mu}(x)$ but with the initial and the final states containing, apart from electrons, positrons, and the photons described by the radiation field $A^{\mu}(x)$, also photons in an appropriate coherent state. Analogously as in the main text, we assume that the background electromagnetic field can be written in Fourier space as 
\begin{equation}
\mathcal{A}^{\mu}(x)=\sum_{j=1}^2\int(d^3\bm{k})[b_j(\bm{k})e^{-i(kx)}e^{\mu}_j(\bm{k})+b^*_j(\bm{k})e^{i(kx)}e^{\mu\,*}_j(\bm{k})].
\end{equation}
In this case, the mentioned coherent state is given by $\ket{B}=D(b)\ket{0}$, where
\begin{equation}
D(b)=\exp\left\{\sum_{j=1}^2\int(d^3\bm{q})[b_j(\bm{q})a^{\dag}_j(\bm{q})-b^*_j(\bm{q})a_j(\bm{q})]\right\}.
\end{equation}
By using the same notation for the initial and final states as above and by exploiting the property $D^{\dag}(b)A^{\mu}(x)D(b)=A^{\mu}(x)+\mathcal{A}^{\mu}(x)$ already discussed in the main text \cite{Mandel_b_2013,Agarwal_b_2013}, we obtain that
\begin{equation}
\begin{split}
\prescript{}{V}{\braket{f(+\infty)|S_V|i(-\infty)}}^{}_V&=\prescript{}{V}{\braket{f(+\infty)|\mathcal{T}\left(e^{-i\int d^4x\,j^{\mu}_V(\tilde{x})[A_{\mu}(\tilde{x})+\mathcal{A}_{\mu}(\tilde{x})]}\right)|i(-\infty)}}^{}_V\\
&=\prescript{}{V}{\braket{f(+\infty)|D^{\dag}(b)\mathcal{T}\left(e^{-i\int d^4x\,j^{\mu}_V(\tilde{x})A_{\mu}(\tilde{x})}\right)D(b)|i(-\infty)}}^{}_V\\
&=\prescript{}{V}{\braket{F(+\infty)|\mathcal{T}\left(e^{-i\int d^4x\,j^{\mu}_V(\tilde{x})A_{\mu}(\tilde{x})}\right)|I(-\infty)}}^{}_V,
\end{split}
\end{equation}
with $\ket{I(-\infty)}_V=D(b)\ket{i(-\infty)}_V$ and $\ket{F(+\infty)}_V=D(b)\ket{f(+\infty)}_V$, which concludes the proof of the equivalence of the two approaches. Note that the equivalence is shown for the states where all the creation operators describing the photons other than those in the coherent state are on the right of the displacement operator $D(b)$. 

\end{widetext}

%\bibliography{arXiv,Books,Reviews,Papers_Radiation,Papers_RR,Papers_Various,Papers_Squeezing,Homepages}

\end{document}